\documentclass[10pt,conference]{IEEEtran}

\usepackage[utf8]{inputenc}
\usepackage{amsmath}
\usepackage{dsfont}
\usepackage{array}
\usepackage{graphicx}
\usepackage{url}

\ifCLASSOPTIONcompsoc 
\usepackage[caption=false,font=normalsize,
            labelfont=sf,textfont=sf]{subfig}
\else
\usepackage[caption=false,font=footnotesize]{subfig}
\fi

\title{Evolving Structures in Complex Systems}

\author{\IEEEauthorblockN{Hugo Cisneros}
  \IEEEauthorblockA{CIIRC, CTU in Prague$^{1}$ \\
    ENS Paris-Saclay \\ hugo.cisneros@ens-paris-saclay.fr}
  \and \IEEEauthorblockN{Josef Sivic}
  \IEEEauthorblockA{Inria, DI-ENS, PSL$^{2}$\\CIIRC,
    CTU in Prague$^{1}$ \\ josef.sivic@inria.fr} \and
  \IEEEauthorblockN{Tomas Mikolov}
  \IEEEauthorblockA{Facebook AI \\ tmikolov@fb.com}}

\date{June 2019}

\begin{document}
\maketitle

\footnotetext[1]{CIIRC - Czech Institute of Informatics, Robotics and
  Cybernetics, Czech Technical University in Prague.}
\footnotetext[2]{WILLOW project, Departement d'Informatique de l'\'Ecole Normale
  Sup\'erieure, ENS/INRIA/CNRS UMR 8548, PSL Research University.}

\begin{abstract}
  In this paper we propose an approach for measuring growth of complexity of
  emerging patterns in complex systems such as cellular automata. We discuss
  several ways how a metric for measuring the complexity growth can be defined.
  This includes approaches based on compression algorithms and artificial neural
  networks. We believe such a metric can be useful for designing systems that
  could exhibit open-ended evolution, which itself might be a prerequisite for
  development of general artificial intelligence. We conduct experiments on 1D
  and 2D grid worlds and demonstrate that using the proposed metric we can
  automatically construct computational models with emerging properties similar
  to those found in the Conway's Game of Life, as well as many other emergent
  phenomena. Interestingly, some of the patterns we observe resemble forms of
  artificial life. Our metric of structural complexity growth can be applied to
  a wide range of complex systems, as it is not limited to cellular automata.
\end{abstract}

\section{Introduction}
Recent advances in machine learning and deep learning have had successes at
reproducing some very complex feats traditionally thought to be only achievable
by living beings. However, making these systems adaptable and capable of
developing and evolving on their own remains a challenge that might be crucial
for eventually developing AI with general learning capabilities (for example as
is further discussed in \cite{mikolov2016roadmap}). Building systems that mimic
some key aspects of the behavior of existing intelligent organisms (such as the
ability to evolve, improve, adapt, etc.) might represent a promising path.
Intelligent organisms --- e.g., human beings but also most living organisms if
we consider a broad definition of intelligence --- are a form of spontaneously
occurring, ever evolving complex systems that exhibit these kinds of properties
\cite{booker_perspectives_2004}. The ability to sustain open-ended evolution
appears to be a requirement in order to enable emergence of arbitrarily complex
adaptive systems.

Although a rigorous attempt at defining intelligence or life is beyond the scope
of this paper, we assume that a system we might identify as evolving, with the
potential of developing intelligence, should have the property of
self-preservation and the ability to grow in complexity over time. These
properties can be observed in living organisms \cite{booker_perspectives_2004}
and should also be a part of computational models that aim to mimic them.

To recognize self-preservation and growth in complexity, one should be able to
detect emerging macro-structures composed of smaller elementary components. For
the purpose of obtaining computational models that grow in complexity over time,
one should also be able to determine the amount of complexity these systems
contain. We propose and discuss in this paper several ways of estimating the
complexity and detecting the presence of emerging and stable patterns in complex
systems such as cellular automata. We show that such metrics are useful when
searching the space of cellular automata with the objective of finding those
that seem to evolve in time.

\section{Related work}

\subsection{Artificial life and open-ended evolution}

Several works have attempted to artificially create open-ended evolution. A
non-exhaustive list of some well known systems include Tierra
\cite{s_ray_approach_1991}, Sims \cite{sims_evolving_1994}, Avida
\cite{ofria_avida:_2004}, Polyworld \cite{yaeger_computational_1994}, Geb
\cite{channon_improving_2003}, Division Blocks \cite{spector_division_2007} and
Chromaria \cite{soros_identifying_2014}. Designs focusing on an objective, and
making use of reinforcement learning methods to drive evolution are also being
studied, e.g. in \cite{pathak_learning_2019}. Most of these simulated ``worlds''
have had some success in reproducing key aspects of evolving artificial life,
enabling the emergence of complex behavior from simple organisms. However, they
still work within constrained simulated environments and usually consider
organisms composed of elementary building blocks, while they don't work outside
of this usually very constrained framework. Divergent and creative evolutionary
process could be happening at a much lower conceptual level with fewer
assumptions. For this reason, we consider cellular automata in the rest of the
paper because they rely on a very few assumptions while offering a very large
expressive power and a potentially wide range of behaviors that can be
discovered. However, the metrics defined in this paper have the potential to be
applied to other types of complex systems as discussed in
Section~\ref{sec:conclusion}.

\subsection{Cellular automata}

Cellular automata are very simple systems, usually defined in one or two
dimensions, composed of cells that can be in a set of states. The cells are
updated in discrete time steps using a transition table that defines the next
state of a cell given the states of its neighbors.
They were originally proposed by Stanislaw Ulam and studied by Von Neumann
\cite{von_neumann_theory_1966}, who was interested in designing a computational
system that can self-reproduce itself in a non-trivial way. The trivial
self-reproducing patterns were then those that do not have potential to evolve,
for example the growth of crystals.

Stephen Wolfram later took a more bottom up approach, beginning with the study
of the simple 1D binary cellular automata (CA), and identifying four qualitative
classes of cellular automaton behavior \cite{wolfram_universality_1984}:
\begin{description}[\IEEEsetlabelwidth{Class 0}]
\item[Class 1]  evolves to a homogeneous state.
\item[Class 2]  evolves to simple periodic patterns.
\item[Class 3]  yields aperiodic disordered patterns.
\item[Class 4]  yields complex aperiodic and localized structures, including
  propagating structures.
\end{description}
Wolfram and his colleagues also studied 2D CA using tools from information
theory and dynamical systems theory, describing the global properties of
these systems in terms of entropies and
Lyapunov exponents \cite{packard_two-dimensional_1985}.

Christopher Langton and colleagues also studied CA dynamics
\cite{li_transition_1990} --- e.g. using the $\lambda$ parameter
\cite{langton_computation_1990} --- and designed a self-replicating pattern much
simpler than Von Neumann's \cite{langton_self-reproduction_1984}, now known as
Langton's loops. The main issues with his system and Von Neumann's universal
replicator is the fact that they are very fragile and based on a large amount of
human design. As a consequence, although they do self-replicate, they cannot
increase in complexity and are not robust to perturbations or unexpected
interactions with the environment.

A genetic algorithm-based search for spontaneously occurring self-replicating
patterns in 2D cellular automata with several states was undertaken in
\cite{bilotta_artificial_2011} using entropy measures of the frequency
distribution of $3\times 3$ patterns.

\subsection{Compression and complexity}
Compression has often been used as a measure of complexity. Lempel and Ziv have
introduced in \cite{lempel_complexity_1976} the now widespread Lempel-Ziv (LZ)
algorithm as a method for measuring the complexity of a sequence. By
constructing back-references to previous parts of a string, the LZ algorithm is
capable of taking advantage of duplicate patterns in the input to reduce its
size. The DEFLATE algorithm that we use in the following section combines LZ
with Huffman coding for efficient representation of the symbols obtained after
the first step. It is one of the most widespread compression algorithms and is
for instance used in gzip and PNG file compression standards.

The PAQ compression algorithm series \cite{mahoney_fast_2000} is an ensemble of
compression algorithms initially developed by Matt Mahoney with state of the art
compression ratio on several compression benchmarks. Better compression of an
input means a better approximation of the minimum description length and
implicit understanding of more of the underlying patterns in input data. The
usefulness of a better compressor is that it can detect much more complex and
intricate patterns that aren't simple repetitions of previous patterns.

In \cite{zenil_compression-based_2010}, H. Zenil investigates the effects of a
compression-based metric to classify cellular automata and observes that it
results in a partitioning of the space of 1D CA into several clusters that match
Wolfram's classes of automata. He also used this approach on the output of
simple Turing machines and some 1D automata with more than two states and larger
neighborhoods. Extensions of this work include asymptotic sensitivity analysis
of the compressed length for input configurations of growing complexity, as
introduced in \cite{zenil_asymptotic_2013, zenil_what_2014}.

Additionally, image decompression time as an approximation of Bennet's logical
depth \cite{bennett_logical_1995, zenil_image_2012} and the output distribution
of simple Turing machines combined with block decomposition of CA to approximate
their algorithmic complexity have also been investigated
\cite{zenil_two-dimensional_2015, soler-toscano_calculating_2014}. However, the
possible extent to which such measures of complexity could be applied to more
complex automata and other complex systems has not yet been extensively studied.
For a review of several measures of complexity and their applications, see
\cite{grassberger_randomness_1989}.

\section{Compression-based metric}\label{sec:compr-based-metr}

A cellular automaton of size $n$ in 1D can be represented at time $t$ by its
grid-state $S^{(t)} = \{c_1^{(t)}, ..., c_n^{(t)}\}$ where each $c_i$ (also
called a cell) can take one of $k$ possible values (representing the possible
states), and a transition rule $\phi$. The transition rule is defined with
respect to a neighborhood radius $r$ with the mapping $\phi(c^{(t)}_{i-r}, ...,
c^{(t)}_i ..., c^{(t)}_{i+r} ) = c^{(t+1)}_i$ that maps $\{1, ..., k\}^{2r+1}$
to $\{1, ..., k\}$. The quantity $2r + 1$ corresponds to the number of
cells taken into account for computing the next state of a cell, namely that
cell itself and $r$ neighboring cells in both directions.

Simulating a CA amounts to the recursive application of this mapping $\phi$ to
an initial state $S^{(0)} = \{c_1^{(0)}, ..., c_n^{(0)}\}$.

In the rest of the paper, we consider cyclic boundary conditions for the
automata, meaning that the indices $i-r, ..., i+r$ above are taken modulo $n$
the size of the automaton in 1D.  Boundary conditions can have some
effect on a CA's evolution, but cyclic boundaries have been empirically
observed to have limited effect on the complexity of automata in 1D
\cite{luvalle_effects_2019}.

The definition given in the equation above can be extended to higher dimensional
automata by modifying the neighborhood and the definition of $\phi$. A 2D
neighborhood of radius 1 can be defined as the 3 by 3 square around the center
cell --- also called the Moore neighborhood --- or by only considering the four
immediate horizontal and vertical neighbors of the center cell --- the Von
Neumann neighborhood.

Elementary cellular automata (ECA) are 1D CA with $k = 2$ and $r = 1$. There are
$2^3$ elements in $\{1, ..., k\}^{2r+1}$ and $2^{2^3} = 256$ possible different
set transition rules that are often compactly represented as a binary string
with 8 bits. The relatively low number of rules of this type makes it possible
to appreciate the performance of a metric and compare it with others.

We define the compressed length $C$ of a 1D cellular automaton at time $t$ as
\begin{align}
  \textstyle
  C(S^{T}) = \text{length}\left(\texttt{comp}(c_1\ ||\ c_2\ ||\ ...\ c_n)\right)
\end{align}

where $||$ denotes the string concatenation operator and the cells $c_i$ are
implicitly converted into string characters (with one symbol per unique state).
\texttt{comp} is a compression algorithm that takes a string as input and
outputs a compressed string, and length is the operator that returns the length
of an input string.

\begin{figure}[tbp]
  \centering \subfloat[6 highest scoring automata. Only the first 30 timesteps
  are shown for readability.]{
    \includegraphics[width=.42\linewidth]{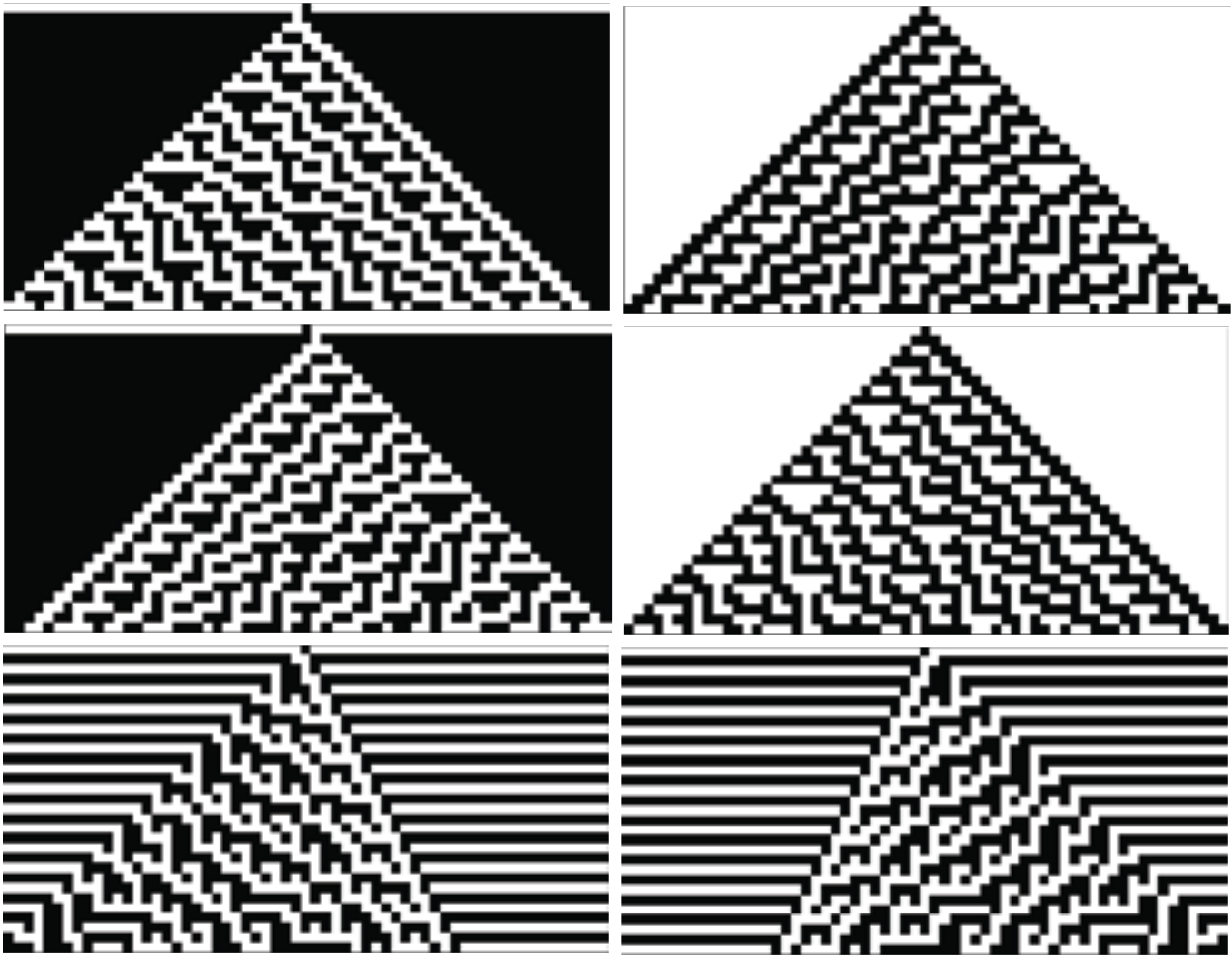}
    \label{subfig:highest_6}
  }
  \hfil
  \subfloat[All 256 compressed length scores]{
    \includegraphics[width=.42\linewidth]{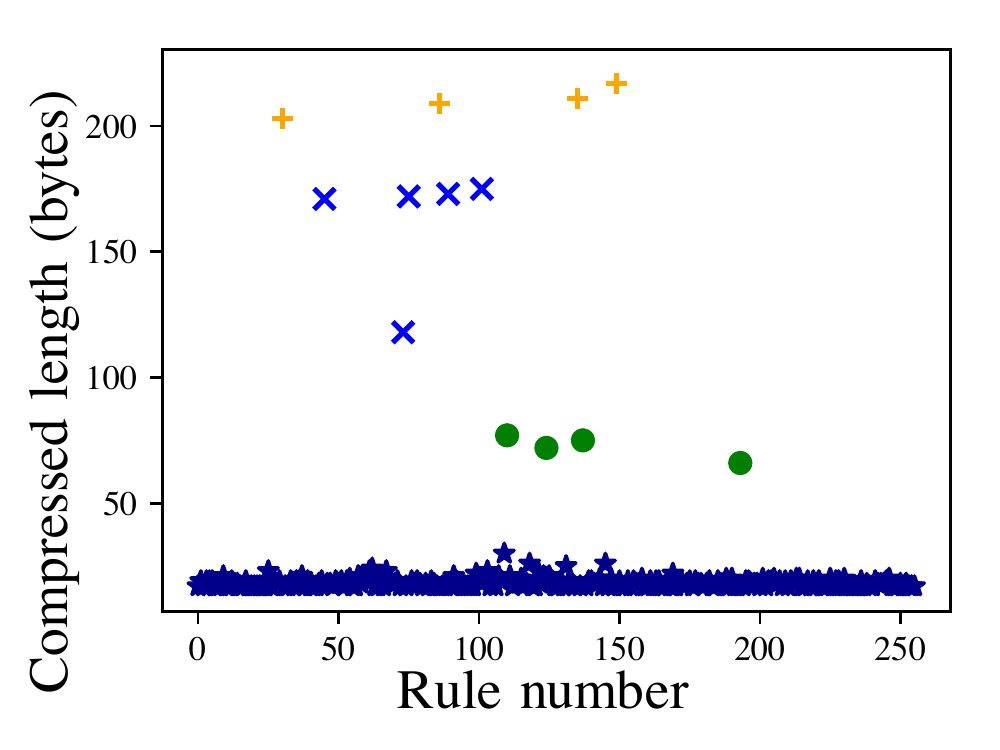}
    \label{subfig:comp_scores}
  }
  \caption{Compression-based metric on 1D ECA. \ref{subfig:highest_6} represents
    the 1D ECA evolution with each line being the state of the automaton at a
    given timestep, starting from a single cell set to 1. Cells which are in
    state 1 are represented in black and cells in state 0 are represented in
    white. Time increases downwards. Figure~\ref{subfig:comp_scores} represents
    the compressed length of the 256 ECA rules, with different marker and colors
    corresponding to the obtained KMeans clusters.}
  \label{fig:comp_eca}
  \vspace{-20pt}
\end{figure}

Similarly to \cite{kowaliw_measures_2008, zenil_compression-based_2010}, we use
zlib’s C implementation of DEFLATE to compress the final state of the automaton.
If we apply the above metric to the 256 ECA run for 512 timesteps and
initialized with one activated cell in the middle, we obtain the plot of
Figure~\ref{subfig:comp_scores}. This example is re-used in the paper as a way
to easily visualize and check that the defined complexity measures are coherent
with one another. The colors on Figure~\ref{subfig:comp_scores} were obtained
with a KMeans clustering algorithm applied on the compressed length of the
automata states.

As visible on Figure~\ref{subfig:comp_scores}, rules are clearly separated into
several clusters that turn out to match Wolfram’s classification of ECA. Class 3
behavior can be found at the top of the plot (highest compressed length, orange
and blue clusters), Class 1 and 2 are clearly separated at the bottom part (not
detailed here) and Class 4 rules (colored in green) lie in between the other
types of behavior. The 6 highest scoring rules are shown on
Figure~\ref{subfig:highest_6} and correspond to Class 3 behavior in Wolfram's
classification. Among the classes of behavior, some sub-clusters can be found
that correspond to similarly behaving rules.

Ultimately, the theoretical goal of using compression algorithms is to approach
the theoretical minimum description length of the input
\cite{grunwald_minimum_2007}. For very regular inputs, this length should be
relatively small and inversely for random inputs. However, gzip and PAQ are
crude approximations of the minimum description length, and may only approach it
in a given context. As an example, compressing text data (a task often performed
with gzip in practice) is much more efficient with a language model that can
assign a very low probability to non grammatically correct sentences. The
Kolmogorov complexity \cite{kolmogorov_three_1968} of a cellular automaton is
upper bounded by a value that is independent of the chosen rule, as it is
entirely determined by the transition table, the grid size, initial
configuration and number of steps.

\section{Predictor-based metric}

One obvious limit of using compression length as a proxy for complexity is the
fact that interesting systems mostly have intermediate compressed length.
Compressed length increases with the amount of disorder in the string being
compressed. Therefore, extreme lengths correspond either to systems that do not
increase in complexity on the lower end of the spectrum, or systems that produce
a maximal amount of disorder on the higher end. Neither of them have the
potential to create interesting behavior and increase in complexity.
Intermediate values of compressed length are also hard to interpret, since
average lengths might either correspond to interesting rules or slowly growing
disordered systems.

To cope with this limitation, one should also take into account the dynamics of
complexity, that is how the system builds on its complexity at a given time as
it keeps evolving, while retaining some of the structures it had acquired
earlier. Compression leverages the amount of repetitions in inputs to further
compress and this may also be used as an estimate of structure overlap, as
explained in the following section.

\subsection{Joint compression}\label{sec:joint-compression}

As a way to both measure the complexity and the amount of overlap between two
automata states apart in time, we define a joint compressed length metric for a
delay $\tau$ as
\begin{align}
  \textstyle C'\left(S^{(T + \tau)}, S^{(T)}\right) =
  C\left(S^{(T)}\ ||\ S^{(T + \tau)}\right)
\end{align}
where $||$ represents the concatenation operator. This quantity is simply the
compressed length of a pair of global states --- defined at the beginning of
\ref{sec:compr-based-metr}, represented by the letter $S$ --- at two timesteps
separated by a delay $\tau$. In 1D, concatenation means chaining the two string
representations before compressing, and in 2D we can chain two flattened
representations of the 2D grid. This introduces several issues which we discuss
in Section~\ref{sec:count-based-pred}.

To quantify the amount of overlap between the two global states, we can compute
the ratio of this joint compressed length with the sum of the two compressed
lengths $C(S^t)$ and $C(S^{t-\tau})$, thereby forming the joint compression
score
\begin{align}
  \textstyle \mu = \dfrac{C\left( S^t \right) +
  C\left( S^{t - \tau} \right)}{C'\left( S^t, S^{t - \tau} \right)}
\end{align}
defined for an automaton $S$, time $t$ and delay $\tau$.

\begin{figure}[htbp]
  \centering
  \includegraphics[width=.8\linewidth]{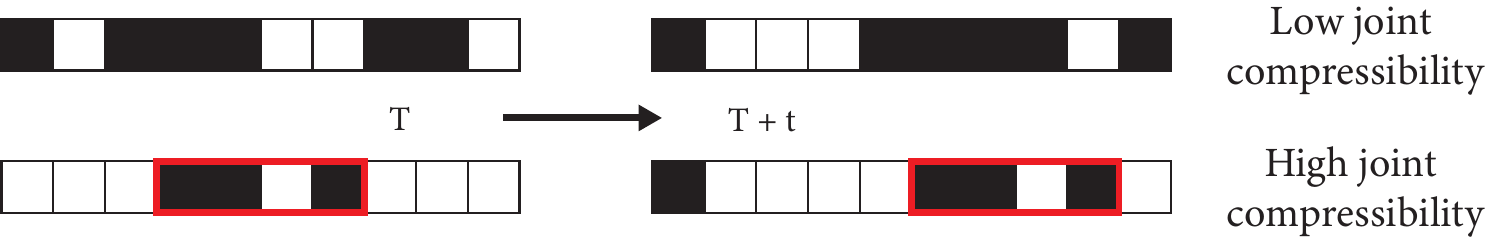}
  \caption{Joint compression method illustration. If a structure persists
    through time, this will decrease the joint compressed length compared to the
    sum of compressed lengths. A persistent structure is circled in red.}
  \label{fig:joint_schema}
  \vspace{-10pt}
\end{figure}

This metric is based on the intuition that if patterns occur at step $T - \tau$
of the automaton's evolution and are still present at step $T$, the joint
compressed length will be lower than the sum of the two compressed length. The
idea is illustrated in Figure~\ref{fig:joint_schema}, where it is pointed out
that a stable moving structure (sometimes called \emph{glider} or
\emph{spaceships} in Game of Life) will yield lower joint compressed lengths.
This also applies to structures that self-replicate, grow from a stable seed or
maintain the presence of some sub-structures. Bigger structures yield a higher
compression gain.

\begin{figure}[htbp]
  \centering
  \subfloat[Highest joint compression score among the 256 ECA.]{
    \includegraphics[width=.4\linewidth]{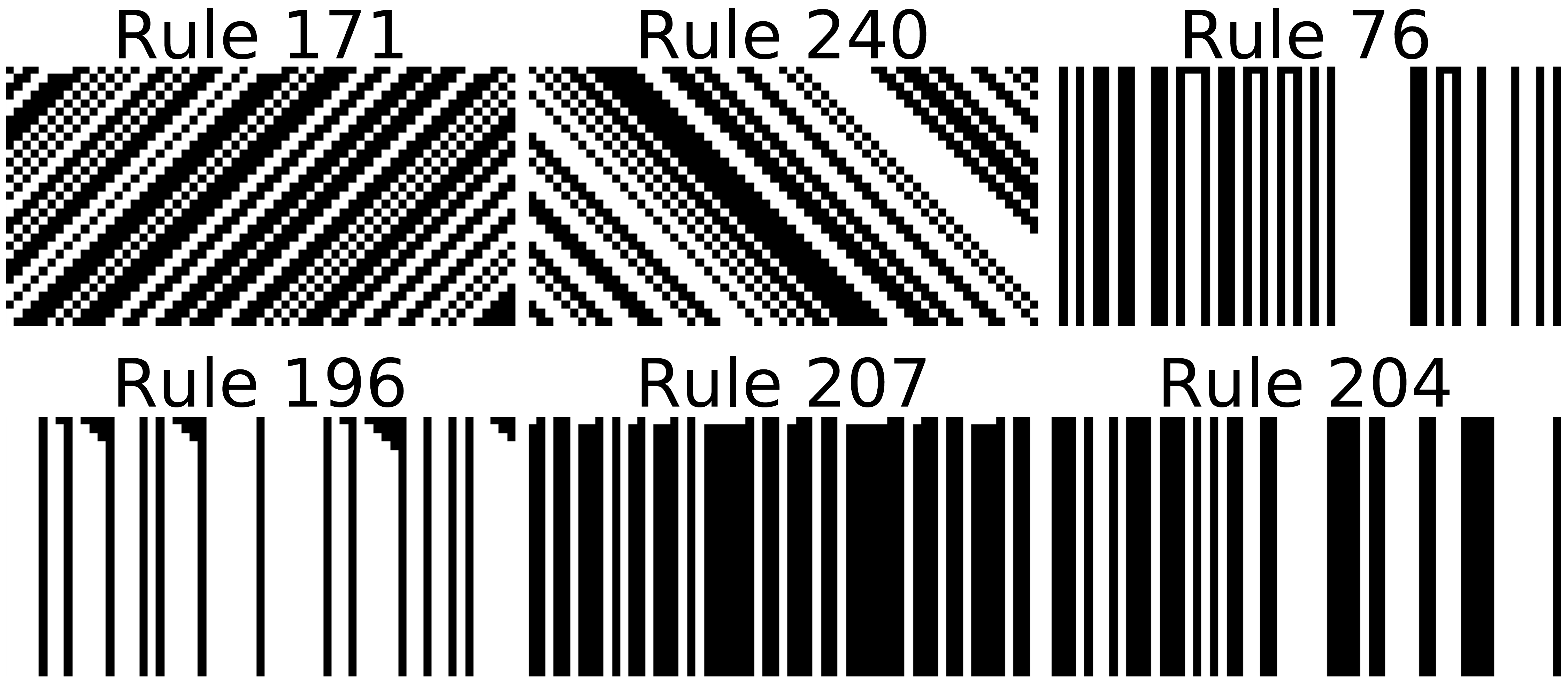}
    \label{fig:high_eca_joint}
  }
  \hfil
  \subfloat[With condition on compressed length increase.]{
    \includegraphics[width=.4\linewidth]{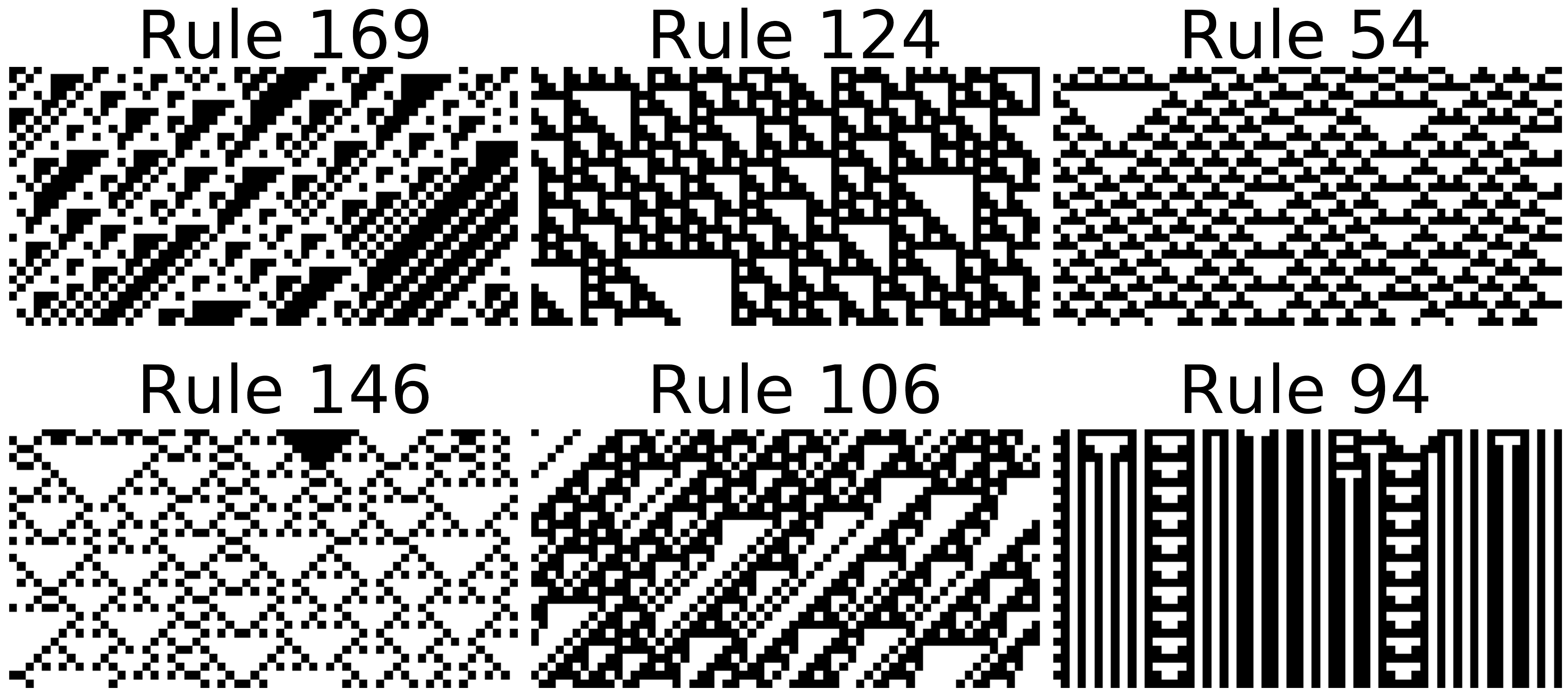}
    \label{fig:high_eca_joint+comp}
  }
  \caption{Comparison of the raw joint compression score and the addition of a
    complexity increase condition. The high overlap in structures is not enough
    to get interesting rules a shown in \ref{fig:high_eca_joint}, but the
    addition of a complexity threshold allows to retrieve rules with complex but
    still structured behavior, as shown in \ref{fig:high_eca_joint+comp}.
    Figures are from the same slice of 60 cells over 30 timesteps taken from
    larger automata with random initial states. The top row corresponds to $t =
    0$ and time increases downwards.}
  \label{fig:joint_highest}
  \vspace{-10pt}
\end{figure}

Joint compression alone is not sufficient since it only selects rules that
either behave like identity or shift input because they maximize the
conservation of structures through time --- as illustrated in
Figure~\ref{fig:high_eca_joint}. However, one may combine the joint compression
score with another complexity measure to only select rules that exhibit some
disorder, or growth in complexity --- as Figure~\ref{fig:high_eca_joint+comp}
shows (the condition here was a threshold on the difference of compressed length
between initial and final states).

\subsection{Count-based predictor}\label{sec:count-based-pred}

A major issue with the joint compression metric is the fact that it is designed
to compress a linear stream of data. This is not ideal when considering higher
dimensional automata. Larger sets of transformations have to be
considered such as translations, rotations, flips, etc. Theoretically this
should not be a problem for a good enough linear compression algorithm, but
hardware and software limitations make it impractical to work with existing
algorithms on higher dimensional structures --- with e.g DEFLATE's upper limit
on dictionary size.

These higher dimensional automata might be better at generating complex
dynamics, and the large size of their rule spaces makes it a challenge to
explore. There has been at least one attempt to deal with these higher
dimensional systems \cite{zenil_two-dimensional_2015} that lacks the scalability
to work with large inputs.

An alternative to the linear compression-based method presented above would be
to use compressors optimized for n-dimensional data (e.g. PNG compression for 2D
automata) to take advantage of spatial correlation for compressing. However,
these compressors are rare for higher dimensional data, and are usually
optimized for one type of input --- e.g. images with PNG.

Another way to tackle the problem is to use a prediction based approach to
compression. Similarly to methods described in
\cite{schmidhuber_sequential_1996} and one of the first steps of the PAQ
compression algorithm \cite{mahoney_fast_2000}, we learn a statistical model of
input data to predict the content of a cell given its immediate predecessors.
For compression, this is often followed by an encoding step --- using Huffman or
arithmetic coding --- that encodes data which contains the least information
(least ``surprising'' data) with the most compact representation. This approach
can also be related to the texture synthesis method described in
\cite{efros_texture_1999}, where the authors learn a non parametric model to
predict the next pixel of a texture given a previously synthesized neighborhood.
Additionally, because we don't need the operation to be reversible as in regular
compression, it is not necessary to limit the prediction model to making
prediction with predecessors only.

For a global state $S = (c_{1}, ... c_i, ..., c_{n})$, the neighborhood of cell
$i$ with radius $r$, denoted $n_{r,i}$ is defined as the tuple $n_{r,i} =
(c_{i-r}, ... c_{i-1}, c_{i+1} ..., c_{i+r})$ --- without the middle cell. The
goal of this method is to estimate the conditional probability distribution $p(s
| n_r)$ of the middle states at timestep $T$ given a neighborhood of radius $r$.
Assuming cell states given their neighborhood can be modeled by mutually
independent random variables, the log-probability of global state $S^{(T)}$ is
written
\begin{align}
  \textstyle \log p(S^{(T)}) = \log \prod_{i=1}^N p(c_i | n_{r,i})  =
  \sum_{i=1}^N \log  p(c_i | n_{r,i})
\end{align}

If the automaton has a very ordered behavior, a model will predict with high
confidence the state of the middle cell given a particular neighborhood. On the
other hand, in the presence of maximal disorder, the middle cell will have an
equal probability of being in every state no matter the neighborhood. In the
latter case, a predictive model minimizing $-\log p(S^{(T)})$ would yield a high
negative log-probability.

A simple possible predictor for such purpose is a large lookup table that maps
all visited neighborhoods to a probability distribution over the states that the
middle cell can be in. State distributions for each neighborhood are obtained by
measuring the frequency of cell states given some observed neighborhoods. We
denote by $\Lambda$ this lookup table, defined for a window of radius $r$, which
maps all possible neighborhoods of size $2r + 1$ (ignoring the middle cell) to a
set of probabilities $p$ over the possible states $\{s_1, ..., s_n\}$, and $p$
can be written $[p_{s_1}, p_{s_2}, ... , p_{s_n}]$. $\Lambda$ is defined by
\begin{equation}
  \begin{aligned}
    \textstyle \Lambda :&& \{s_1, ..., s_n\}^{2r} &\to&& \Delta_n\\
    && n_{r,i} &\mapsto&& p
  \end{aligned}
\end{equation}
where $\Delta_n$ denotes the probability simplex in dimension $n$.

\begin{figure}[htbp]
  \centering
  \includegraphics[width=.7\linewidth]{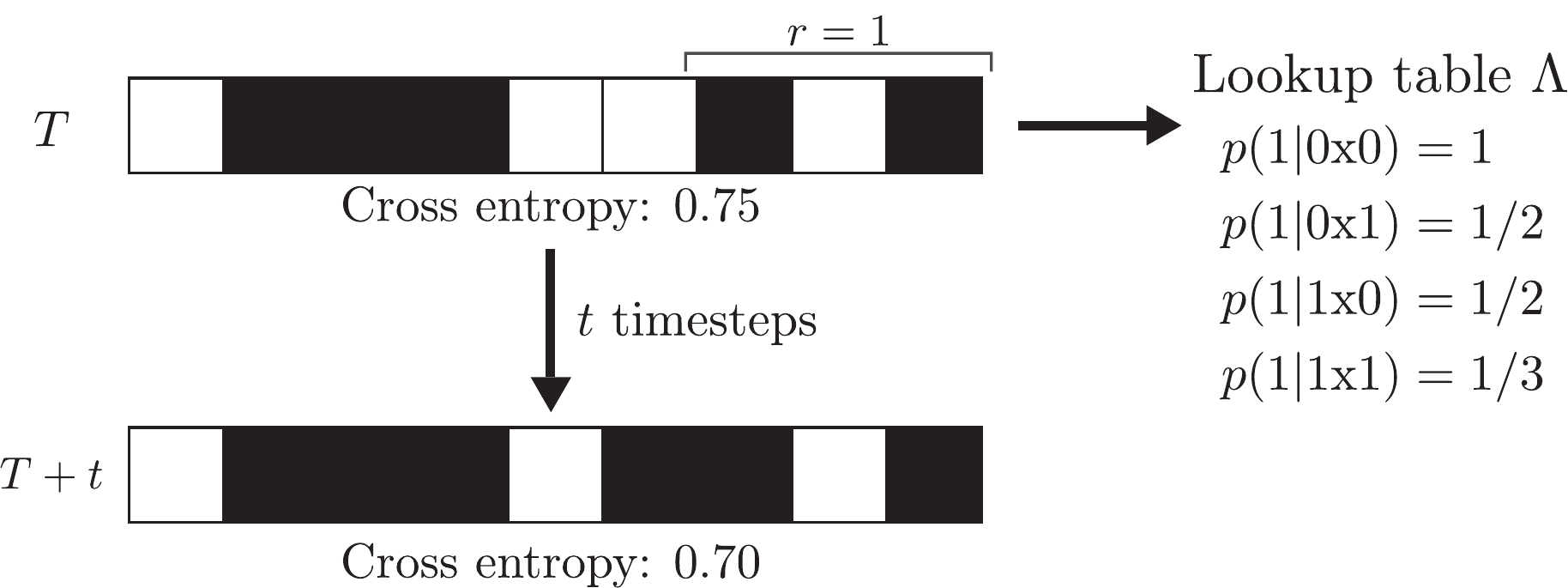}
  \caption{Count-based predictor method for a radius $r=1$. A
    frequency lookup table is computed from the global state at time $T$ by
    considering all neighborhoods with radius $r=1$ (3 consecutive cells
    but ignoring the middle cell). Cross-entropy with the automaton at time $T$
    quantifies the overall complexity. This can be compared to the cross-entropy
    at time $T + t$ for the amount of overlap.}
  \label{fig:schema-count}
  \vspace{-15pt}
\end{figure}

To measure the uncertainty of that predictor, we can compute the cross-entropy
loss between the data distribution it was trained on and its output. We compute
the log probability of the observed data given the model, or the quantity
\begin{align}
  \textstyle
  L = - \frac{1}{N}\sum_{i=1}^N \sum_{k=1}^n \mathds{1}_{\{ s_k \}}(c_i)
  \log\Lambda(n_{r,i})_{s_k}
  \label{eq:loss_count}
\end{align}
where $\mathds{1}_{\{s_k\}}$ denotes the indicator function of the singleton set
$\{s_k\}$. An illustration of the counting process is represented in
Figure~\ref{fig:schema-count}. The quantity $L$ is minimal when the
$\Lambda(n_{r,i})_{s_k}$ always equal one, which means the state of every cell
is entirely determined by its neighborhood.

\begin{figure}[htbp]
  \centering
  \subfloat[Count-based predictor]{
    \includegraphics[width=.405\linewidth]{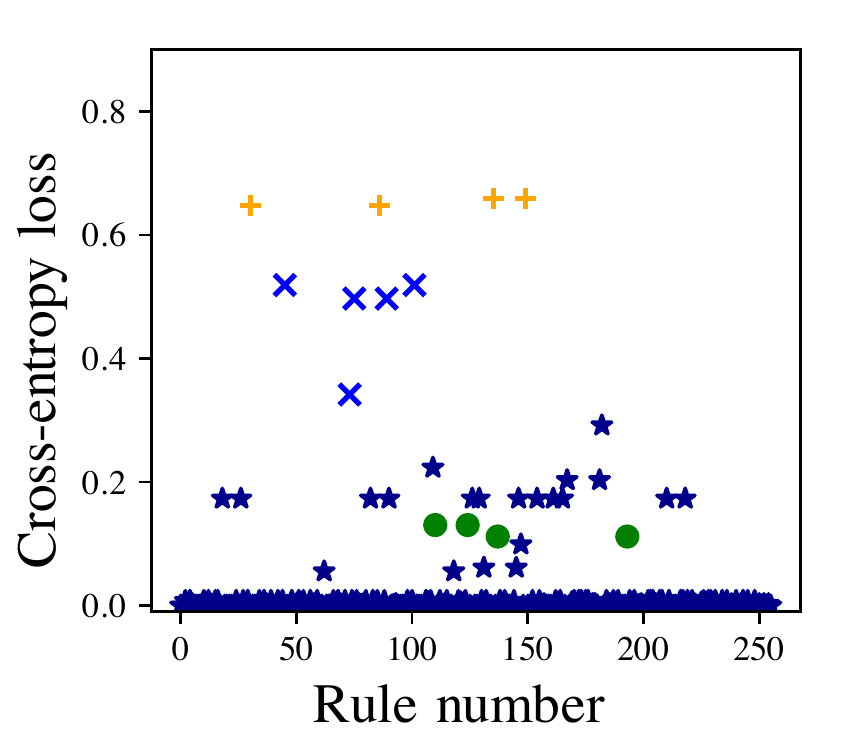}
      \label{subfig:cross_ent_one}
  }\hfil
  \subfloat[Neural network-based predictor]{
    \includegraphics[width=.38\linewidth]{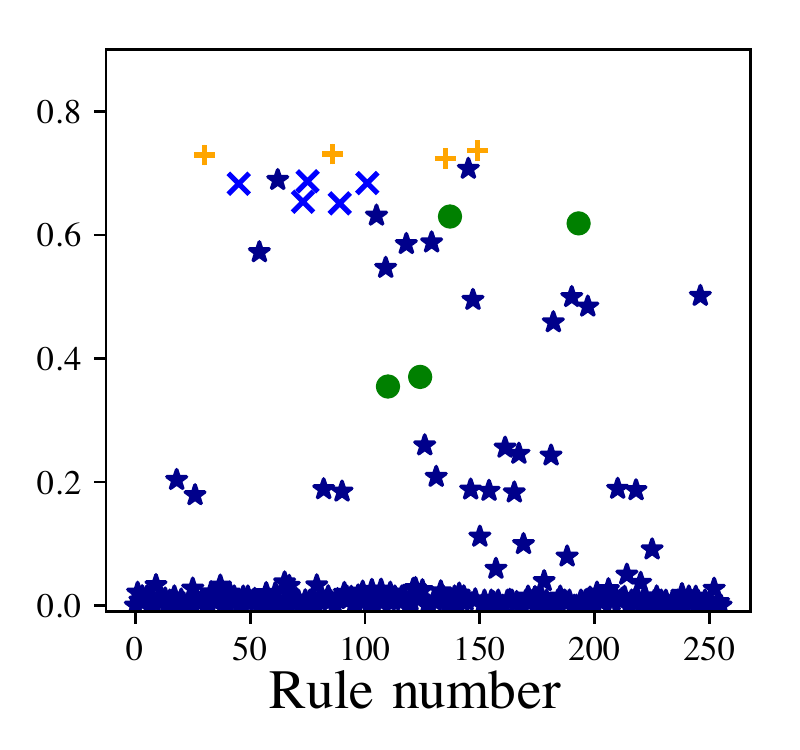}
      \label{subfig:nn_ent_one}
  }
  \caption{Average cross entropy loss for the two predictor-based methods on all
    256 ECA. Rules are separated in several clusters. The
    count-based predictor (left plot) and neural network-based predictor (right
    plot) were applied with a neighborhood radius $r=1$ and $3$.}
\end{figure}

We apply this metric to all 256 ECA, with a window radius of size 3 (the 6
closest neighbors are used for prediction), and the same settings as for
Figure~\ref{subfig:comp_scores}. Cross-entropy loss of the lookup table gives
the results of Figure~\ref{subfig:cross_ent_one}. Colors are the same as in
Figure \ref{subfig:comp_scores} for comparison purposes.

We note the similarity between this plot and the one from Figure
\ref{subfig:comp_scores}, with a roughly equivalent resulting classification of
ECA rules, with the exception of rules with low score. Rules that produce highly
disordered patterns are on top of the plot whereas the very simply behaving
rules are at the bottom. This indicates coherence between the two metrics.

\subsection{Neural network based predictor}

\begin{figure}[htbp]
  \centering
  \includegraphics[width=.7\linewidth]{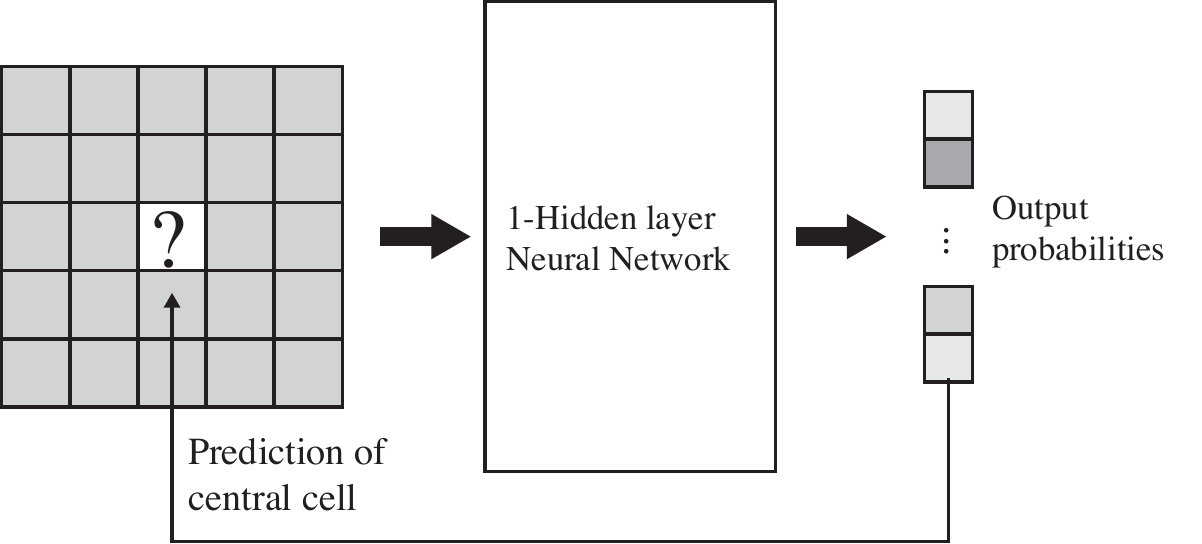}
  \caption{Neural network architecture for predicting a central cell given its
    neighbors. Output probabilities are defined for all possible states of the
    central cell.}
  \label{fig:nn_archi}
\end{figure}

The frequency based predictor described above still has limitations:
\begin{itemize}
\item It doesn't take into account any redundancy in the input which may lead to
  suboptimal predictions (in a CA, very similar positions might have similar
  center cell state distribution, e.g. a glider in Game of Life should be
  recognized by the model no matter the rest of the neighborhood).
\item For the same reasons, when considering large window sizes, the number of
  possible neighborhood configuration gets much larger than the number of
  observed ones, leading to an input sparsity problem.
\end{itemize}
More sophisticated models can cope with above limitations by dealing with high
dimensional inputs without sparsity problems, and taking into account redundancy
of inputs and potential interactions between states for prediction.

We measure the cross-entropy loss of this simple model on the training set after
a standard learning procedure which is the same for all rules. The procedure is
applied to a one hidden layer neural network with a fixed hidden layer size. We
use a ReLU non-linearity for the hidden layer and a softmax to obtain the output
probabilities.

For $n$ possible states ${s_1, ..., s_n}$, a cell in state $s_k$ is represented
as a vector of 0s of size $n$ with a 1 in position $k$. The input to the network
is the concatenation of these cell vectors for all cells in the neighborhood.
The output of the network is a vector of size $n$ with the output probability
for each state.

Gradient updates are computed during training to minimize the cross-entropy loss
between outputs and target examples. For a timestep $T$, we use the training
procedure in order to minimize with respect to $\theta$ the following quantity,
\begin{align}
  L_\theta^{(T)} = - \frac{1}{N}\sum_{i=1}^N \sum_{k=1}^n
  \mathds{1}_{\{ s_k \}}\left(c_i^{(T)}\right)
  \log\left[f_\theta\left(n_{r,i}^{(T)}\right)_{s_k}\right]
  \label{eq:loss_network}
\end{align}
where the neural network depending on parameter $\theta$ is denoted $f_\theta$,
and $n_{r,i}^{(T)}$, the neighborhood of cell $i$ with radius $r$ at time $T$ is
defined in the same way as in eq.~\eqref{eq:loss_count}. Loss is computed with
respect to the testing set at time $T + \tau$ by computing the same quantity at
this subsequent timestep.

The training procedure is selected to achieve reasonable convergence of the loss
for the tested examples. It must be well defined and stay the same to allow for
comparison of the results across several rules. Score at timestep $T$ for a
delay $\tau$ is computed with the following formula
\begin{align}
  \textstyle \mu_\tau = \dfrac{L^{(T)}}{L^{(T + \tau)}}
  \label{eq:main_metric}
\end{align}
where $L^{(T + \tau)}$ is the log probability of the automaton state at timestep
$T + \tau$ (defined in eq.~\eqref{eq:loss_network}) according to a model with
parameters learned during training at timestep $T$ and $L^{(T)}$ is the same as
in eq.~\eqref{eq:loss_network}. The value $\mu_\tau$ will be lower for easily
``learnable'' global states that do not translate well in future steps --- they
create more complexity or disorder --- thereby discarding slowly growing very
disordered structures. Higher values of $\mu_\tau$ correspond to automata that
have a disordered global state at time $T$ that can be transposed to future
timesteps relatively well. Those rules will tend to have interesting spatial
properties --- not trivially simple but not completely disordered because the
model transposes well --- as well as a large amount of overlap between a given
step and the future ones, indicating persistence of the spatial properties from
one state to another. We also selected the metric among other quantities
computed from $L^{(T)}$ and $L^{(T+\tau)}$ because it yielded the best score on
our experimental datasets.

\section{Experiments}\label{sec:experiments}

We carried out several experiments on a dataset of 500 randomly generated 3
states ($n=3$) rules with radius $r=1$. Those rules were manually annotated for
interestingness, defined as the presence of visually detectable non trivial
structures. The dataset contains 46 rules labeled as interesting and 454
uninteresting rules. Ranking those rules with the metrics introduced above
allows to study the influence of parameters and the adequacy between
interestingness as we understand it and what the metric measures.

The task of finding interesting rules can either be framed as a classification
problem or a ranking problem with respect to the score we compute on the
dataset. The performance of our metric can be measured with usual evaluation
metrics used on these problems, and notably the average precision (AP) of the
resulting classifier.

Average precision scores for the neural network and count-based methods for time
windows of 5, 50 and 300 timesteps are given in Table~\ref{experiments_table}.
Scores were computed on automata of size $256\times 256$ cells, ran for 1000
timesteps ($T + \tau = 1000$). Scores were computed for radii ranging from 1
cell (8 nearest neighbors) to 5 cells (120 neighbors), with a one layer neural
network containing 10 hidden units trained for 30 epochs with batches
of 8 examples. Best AP for each time window is shown in bold.
Results for the frequency lookup table predictor are only shown for $r=1, 2$
because of sparsity issues with the lookup table from $r=2$ and above making it
unpractical to use the table --- $3^{24}$ possible entries for the lookup table
with $r=2$ against only $256^2$ observed states.

\begin{table}[t!]
  \renewcommand{\arraystretch}{1}
  \caption{Experimental results - AP scores}
  \label{experiments_table}
  \centering
  \begin{tabular}{|c|c|c|c|c|c|}
    \hline
    \bfseries Neural network& $r=1$ &  2  & 3  & 4 & 5\\
    \hline
    5 steps & 0.387 & 0.448 & \bfseries 0.541 & 0.525 & 0.534\\
    50 steps & 0.377 & 0.433 & 0.517 & 0.491 & \bfseries 0.542\\
    300 steps &0.358 & 0.454 & 0.488 & \bfseries 0.527 & 0.525\\
    \hline
    \bfseries Lookup table& $r=1$ &  2  & & &\\
    \hline
    5 steps & 0.092 & 0.070&&&\\
    50 steps & 0.102 & 0.070&&&\\
    300 steps & 0.093 & 0.069&&&\\
    \hline
  \end{tabular}
  \\[+8pt]
  \begin{flushleft}{\footnotesize This table shows the average precision (AP)
      scores obtained on the dataset of section \ref{sec:experiments} with the
      neural network-based and lookup table-based methods. Results are show for
      delays $\tau = 5, 50, 300$ and several radii values $r$.}\end{flushleft}
  \vspace{-20pt}
\end{table}

From these experiments, bigger radii appear to perform slightly better, although
not in a radical way. Since the number of neighbors scales with the square of
the radius, reasonably small radii might be a good trade-off between performance
and computational cost of the metric.

We also study the performance of our metrics --- lookup table and neural
network-based --- as inputs of a binary classifier against two simple baselines
on a random 70/30 split of our dataset. The first baseline classifies all
example as negative. The second baseline is based on compressed length as
defined in \cite{zenil_compression-based_2010} and computed by choosing a pair
of thresholds that minimize mean square error when classifying examples in
between as positive --- this is based on the observation made in
Section~\ref{sec:compr-based-metr} that interesting rules have intermediate
compressed lengths. Results are in Table~\ref{experiments_table2} where only the
best radius is shown. The lookup table performs better than the baselines but
the neural network gives the best score.

\begin{table}[t!]
  \renewcommand{\arraystretch}{1.2}
  \caption{Experimental results - Accuracy}
  \label{experiments_table2}
  \centering
  \begin{tabular}{|l|m{.1\linewidth}|
    m{.25\linewidth}|
    m{.1\linewidth}|
    m{.11\linewidth}|}
    \hline
    \bfseries  Metric & Baseline & Compressed length
    \cite{zenil_compression-based_2010} &
    Lookup Table & Neural network \\ \hline
    \bfseries Accuracy & 0.90 & 0.913 & 0.926 & \bfseries 0.953 \\ \hline
  \end{tabular}
  \\[+8pt]
  \begin{flushleft}{\footnotesize Accuracy of each metric of complexity when
      used to classify which automatons do evolve interestingly, compared
      against the trivial all-negative baseline and the compressed length
      metric~\cite{zenil_compression-based_2010}.}\end{flushleft}
  \vspace{-20pt}
\end{table}

Above experiments demonstrate the capability of our proposed metric to match a
subjective notion of interestingness of our labeling. For instance, the top 5
and top 10 scoring rules of the best performing configuration ($r=3$, $\tau =
5$) are all labeled as interesting, and top 100 scores contain 41 of the 46
rules labeled as interesting.

\section{Discussion}

In this section, we discuss the results obtained by using the metric of equation
\eqref{eq:main_metric} and the way they can be interpreted.

\paragraph{One dimensional cellular automata}
By applying the metric on the same example as before, we again obtain a
plot with a rule classification that matches a visual appreciation of
complexity of 1D CA. Results are shown on Figure~\ref{subfig:nn_ent_one}.
Similarly to the previous cases, rules we might label as interesting are
unlikely to be either at the top or bottom of the plot.

\paragraph{Two dimensional cellular automata}

Simulations conducted with 2D CA used grids of size 256$\times$256. Automata
were ran for 1000 steps (the metric is measured with respect to the reference
time $T = 1000$). Rules are defined with a table of transitions from all
possible neighborhood configurations with radius $r=1$ (3$\times$3 squares) to a
new state for the central cell. Unbiased sampling of rules, obtained by
uniformly sampling the resulting state for each transition independently,
overwhelmingly produces rules with a similar amount of transitions towards each
state and fails to produce rules without completely disordered behavior more
than 99\% of the time.

Therefore, we adopt a biased sampling strategy of the rules, selecting the
proportion of transitions towards each state uniformly on the simplex --- e.g
for 3 states we might get the triple $(0.1, 0.5, 0.4)$ and sample transitions
according to these proportions. This parametrization can be related to Langton's
lambda parameter \cite{langton_computation_1990} that takes into account the
proportion of transitions towards a transient (inactive) state and all the other
states. We obtain approximately 10\% interesting rules with this sampling as the
proportions of our experimental dataset show.

\begin{figure}[t]
  \centering
  \subfloat{
    \includegraphics[width=.38\linewidth]{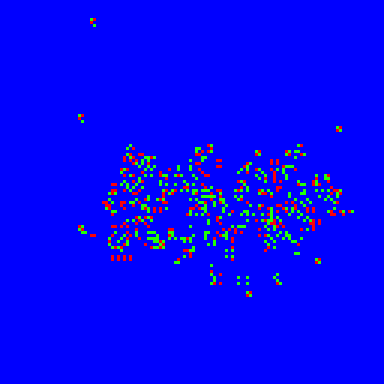}
  }\hfil
  \subfloat{
    \includegraphics[width=.38\linewidth]{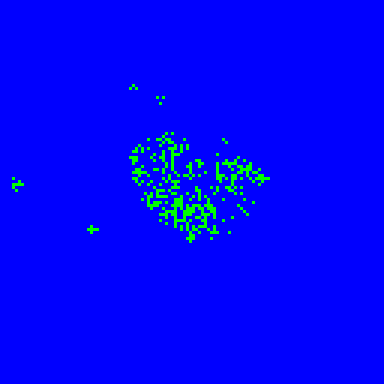}
  }
  \caption{Rules with 3 states that have spontaneously occurring glider
    structures. The gliders are the small structures that are outside of the
    center disordered zone. Some of them move along the diagonals while some
    others follow horizontal or vertical paths. Note that some repeating
    patterns occur also in the more disordered center zone.}
  \label{fig:gliders}
  \vspace{-15pt}
\end{figure}

\begin{figure}[t]
  \centering
  \subfloat{
    \includegraphics[width=.38\linewidth]{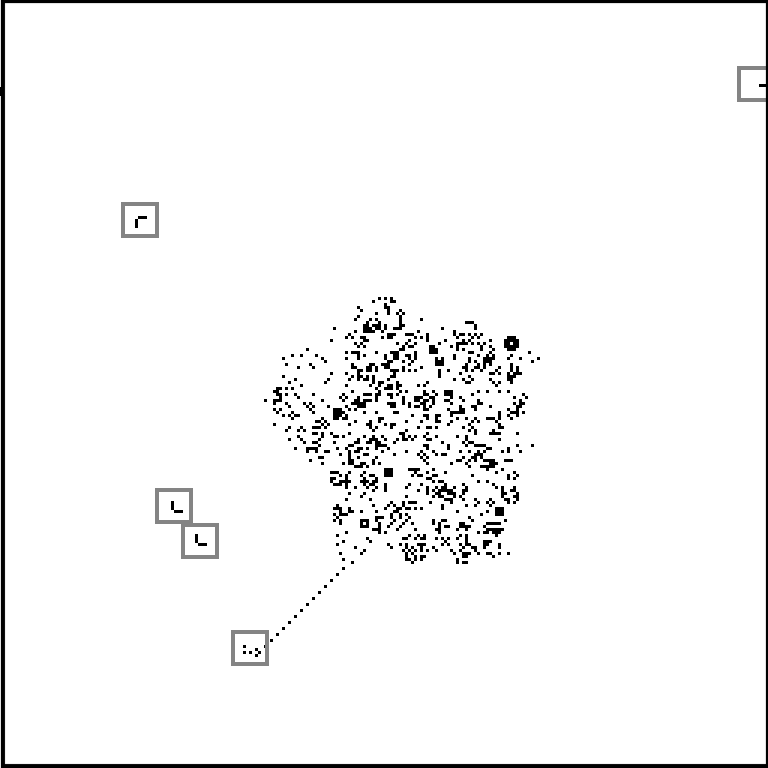}
  }\hfil
  \subfloat{
    \includegraphics[width=.38\linewidth]{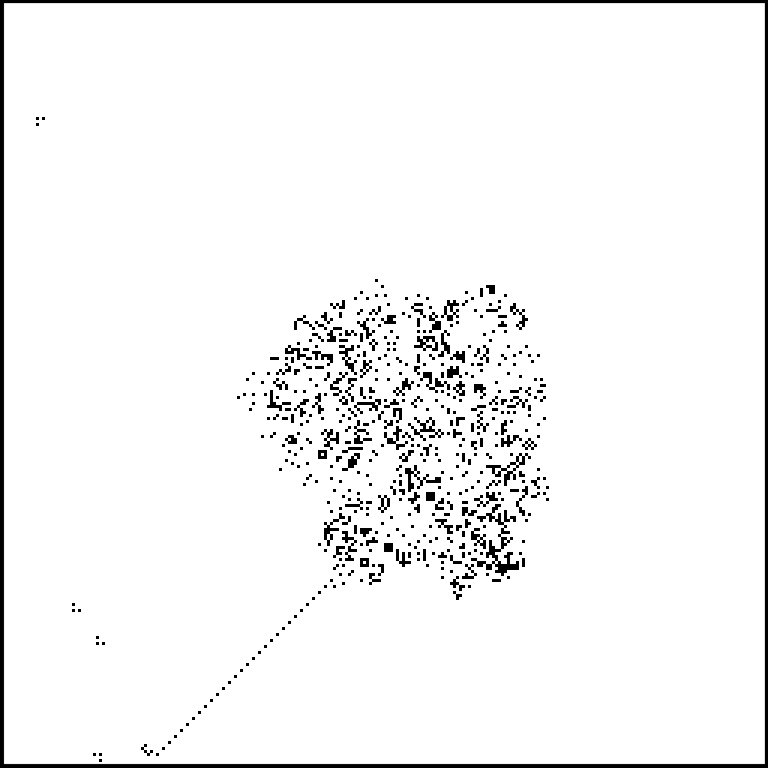}
  }\hfil\setcounter{subfigure}{0} % Get the numbers right
  \subfloat[Timestep $T$]{
    \includegraphics[width=.38\linewidth]{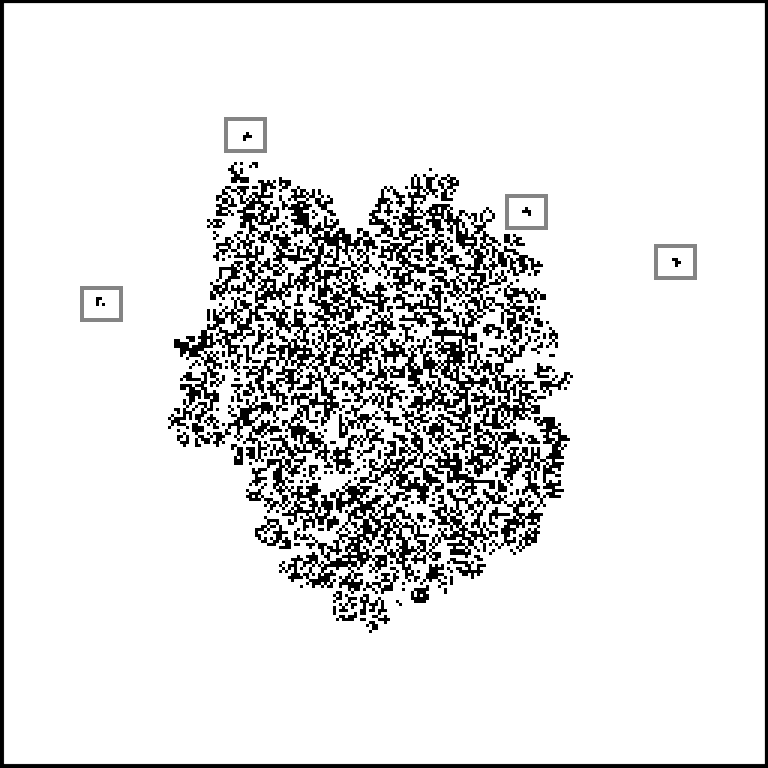}
  }\hfil
  \subfloat[Timestep $T + 50$]{
    \includegraphics[width=.38\linewidth]{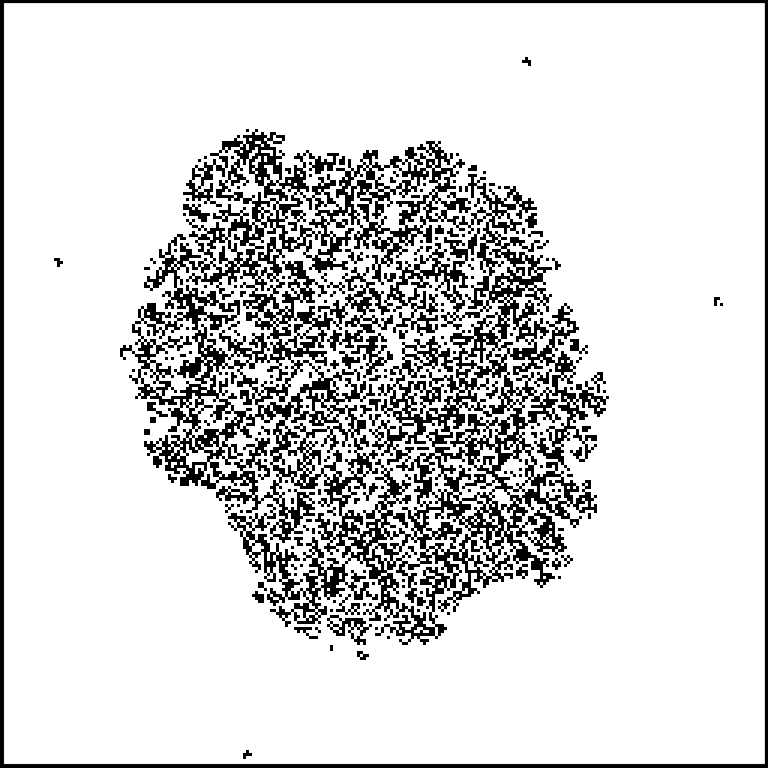}
  }
  \caption{Spontaneous glider formation and evolution is observed for some high
    scoring 2 states rules. Each row corresponds to a rule, with a 50 timesteps
    difference between the two columns. Gliders are marked with a gray square.
    Runs were initialized with a small 20 by 20 disordered square (uniformly
    sampled among possible configuration) in the center simulated for up to 400
    steps.}
  \label{fig:automat_glider}
  \vspace{-15pt}
\end{figure}

Using the neural network-based complexity metric, we were able to find rules
with interesting behavior among a very large set through random sampling. Some
of these rules are shown in the paper. Figure~\ref{fig:automat_glider} displays
three 2D rules that were selected manually upon visual inspection among the 20
highest scoring for metric $\mu_{50}$ (defined in eq.~\eqref{eq:main_metric}) of
a sample of 1700 randomly generated 2-states 3 by 3 neighborhood rules. For
comparison, Conway's Game of Life rule (GoL) ranks in the top 1\% of the 2500
rules mentioned above for runs that don't end in a static global state. We
observe that spontaneous glider formation events appear to be captured by our
metric. Although gliders in cellular automata are a simple process that can
manually be created, detection of their spontaneous emergence within a random
search setting is a first step towards finding more complex macro structures
that can emerge out of simple components. Rules with low scores are
overwhelmingly of the disordered kind.

\begin{figure}[t]
  \centering
  \subfloat{
    \includegraphics[width=.38\linewidth]{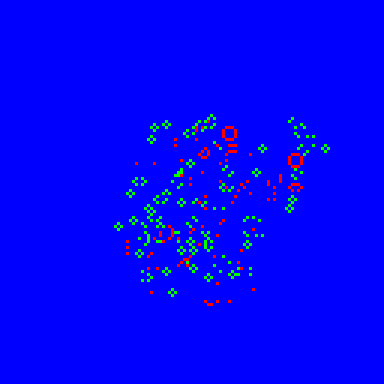}
  }
  \hfil
  \subfloat{
    \includegraphics[width=.38\linewidth]{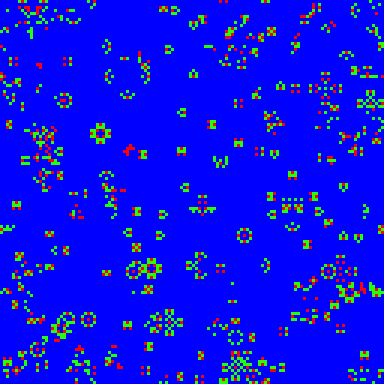}
  }
  \caption{Rules with 3 states that generate cell-like interacting structures.
    These patterns are either static or moving and can interact with one another
    to generate copies of themselves and other patterns. Note the very similar
    micro-structures that are repeated at several places in the space.}
  \label{fig:micro}
  \vspace{-15pt}
\end{figure}

\begin{figure}
  \centering
  \subfloat{
    \includegraphics[width=.38\linewidth]{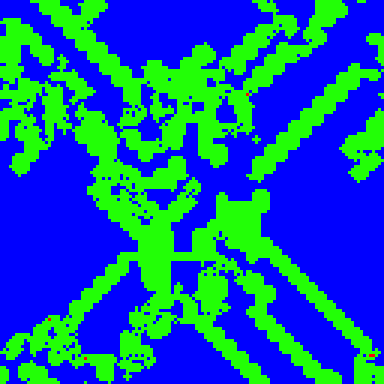}
  }
  \hfil
  \subfloat{
    \includegraphics[width=.38\linewidth]{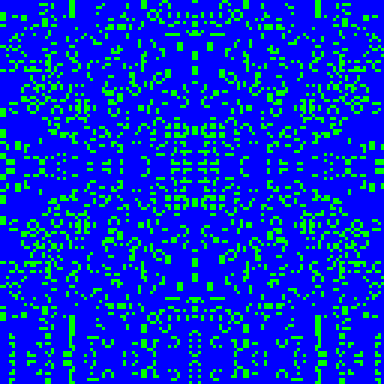}
  }
  \caption{Rules with surprising behaviors that are highly structured but
    complex. Those rules were selected among high-ranking rules for the
    neural-network based complexity metric. They all exhibit structurally non
    trivial behavior.}
  \label{fig:odd}
  \vspace{-10pt}
\end{figure}

Figures~\ref{fig:gliders}, \ref{fig:micro} and \ref{fig:odd} show some three
states rules that were selected through random sampling on the simplex with the
neural-network based metric. They were selected among the 30 highest scoring
rules out of 2500 randomly selected 3 states rules. Their behaviors all involve
the growth and interaction of some small structures made of elementary cells.

All automata were initialized with a random disordered square of 20 by 20 cells
in the center. In the Figures mentioned above, colors were normalized with the
most common state set to blue. Figure~\ref{fig:gliders} shows rules that
spontaneously emit gliders that go through space in a direction until they
interact with some other active part of the automaton. Figure~\ref{fig:micro}
shows rules that generate small structures of between four and thirty cells that
are relatively stable and interact with each other. These elementary components
could be a basis for the spontaneous construction of more complex and bigger
components. Figure~\ref{fig:odd} shows some other rules from this set of high
ranking automata. They highlight the wide range of behaviors that can be
obtained with these systems. Interesting rules from this paper can be found,
along with other examples, in the form of animated
GIFs\footnote{\url{https://bit.ly/interesting_automata}}.

For some of these rules interesting patterns appear less frequently in smaller
grids, indicating that the size of the space might impact the ability to
generate complex macro-structures. Increasing the size of the state space to
very large grids might therefore make it easier generating very complex patterns.

\section{Conclusion}\label{sec:conclusion}

In this paper, we have proposed compression-inspired metrics for measuring a
form of complexity occurring in complex systems. We demonstrated its usefulness
for selecting CA rules that generate interesting emergent structures from very
large sets of possible rules. In higher dimensions where linear compression ---
as in gzip --- is not sufficient to find complex patterns, our metric is also
useful.

We study 2 and 3 states automata in the paper and we plan to investigate the
effects of additional states or larger neighborhoods on the ability to evolve
more structures and obtain more interesting behaviors.

In the future, we will publish the dataset and code to enable reproducibility
and improvement on the results reported
here\footnote{\url{https://github.com/hugcis/evolving-structures-in-complex-systems}}.
The metrics we introduce in this paper could be used to design organized systems
of artificial developing organisms that grow in complexity through an
evolutionary mechanism. A possible path toward such systems could start by
creating an environment where computational resource allocation favors the
fraction of subsystems that perform the best according to our measure of
complexity.

The proposed metric is theoretically applicable to any complex system where a
notion of state of an elementary component and locality can be defined. With
these requirements fulfilled, we can build a similar prediction model that uses
information about local neighbors to predict the state of a component and
thereby assess the structural complexity of an input.

We believe that the capability of creating evolving systems out of such
elementary components and with few assumptions could be a step towards AGI. By
devising ways to guide this evolution in a direction we find useful, we would be
able to find efficient solution to hard problems while retaining adaptability of
the system. It might be suitable to avoid over-specialization that can happen in
systems designed to solve a particular task --- e.g. reinforcement learning
algorithms that can play games, and supervised learning --- by staying away from
any sort of objective function to optimize and by leaving room for open-ended
evolution.

{
\small
\noindent {\bf Acknowledgments.}
This work was partially supported by ERC grant LEAP No.\ 336845, CIFAR Learning
in Machines $\&$ Brains program and the EU Structural and Investment Funds,
Operational Programe Research, Development and Education under the project
IMPACT (reg. no. CZ$.02.1.01/0.0/0.0/15\_003/0000468$). }

\bibliography{biblio}
\end{document}